\documentclass[manuscript]{aastex}
\usepackage{lscape}
\usepackage[]{natbib}
\begin{document}

\title{Quantum Calculation of Inelastic CO Collisions with H. III. Rate Coefficients for Ro-vibrational Transitions}

\author{L.~Song\altaffilmark{ 1}, N.~Balakrishnan\altaffilmark{2}, K.~M.~Walker\altaffilmark{3},
        P.~C.~Stancil\altaffilmark{3}, W.~F.~Thi\altaffilmark{4}, I.~Kamp\altaffilmark{5}
        A.~van~der~Avoird\altaffilmark{1}, G.~C.~Groenenboom\altaffilmark{1} }

\altaffiltext{1}{Theoretical Chemistry,
             Institute for Molecules and Materials,
             Radboud University,
             Heyendaalseweg 135, 6525 AJ Nijmegen,
             The Netherlands}
\altaffiltext{2}{Department of Chemistry, University of Nevada, Las Vegas, NV 89154, USA}
\altaffiltext{3}{Department of Physics and Astronomy and Center for Simulational Physics,
              The University of Georgia, Athens, GA 30602, USA}
\altaffiltext{4}{Max Planck Institute for Extraterrestrial Physics, Garching, Germany}
\altaffiltext{5}{Kapteyn Astronomical Institute, PO Box 800, 9700 AV Groningen, The Netherlands}

\begin{abstract}
We present calculated rate coefficients for ro-vibrational transitions
of CO in collisions with H atoms for a gas temperature range of 10~K
$\leq T \leq$ 3000~K, based on the recent three-dimensional \textit{ab
initio} H-CO interaction potential of \citet{song:13}. Rate coefficients
for ro-vibrational $v=1,j=0-30 \rightarrow v'=0, j'$ transitions were
obtained from scattering cross sections previously computed with the
close-coupling method by \citet{song:15}. Combining these with the rate
coefficients for vibrational $v=1-5 \rightarrow v' < v$ quenching
obtained with the infinite-order sudden approximation, we propose a new
extrapolation scheme that yields the rate coefficients for
ro-vibrational $v=2-5,j=0-30 \rightarrow v',j'$ de-excitation. Cross
sections and rate coefficients for ro-vibrational $v=2, j=0-30
\rightarrow v'=1,j'$ transitions calculated with the close-coupling
method confirm the effectiveness of this extrapolation scheme. Our
calculated and extrapolated rates are very different from those that
have been adopted in the modeling of many astrophysical environments.
The current work provides the most comprehensive and accurate set of
ro-vibrational de-excitation rate coefficients for the astrophysical
modeling of the H-CO collision system. Application of the previously
available and new data sets in astrophysical slab models shows that the
line fluxes typically change by 20-70\% in high temperature environments
(800~K) with an H/H$_2$ ratio of 1; larger changes occur for lower
temperatures.
\end{abstract}


\section{Introduction}
\label{sec:intro}
As the second most abundant molecule in the universe, carbon monoxide
(CO) is commonly detected in a variety of astrophysical environments. As
early as the 1970's, the infrared ro-vibrational bands of CO were
observed in late-type stars \citep{thompson:69, johnson:70, johnson:72}.
Later, \citet{scoville:80} and \citet{ayres:86} detected CO
ro-vibrational lines in the spectra of young stellar objects and the
Sun. In a sample of nine infrared sources identified by
\citet{mitchell:90}, eight sources show evidence of a hot gas component
with temperatures of 120 to 1010~K and CO ro-vibrational emission.
Recently, $\Delta v=1$ ro-vibrational transitions of CO near 4.7 $\mu$m
have been observed in star-forming regions, protoplanetary disks,
and external galaxies using high-resolution spectrometers \citep{
gonzalez:02,najita:03,rettig:04,carmona:05,salyk:07,pontoppidan:08,
brittain:09,vanderplas:09,goto:11,brown:13,bertelsen:14}.
Analysis of these CO spectral lines provides the physical conditions and
chemical composition of the interstellar gas, in particular the kinetic
temperature, column density, volume density, and molecular abundances.
Many of the analyses assumed that the CO ro-vibrational emission is
based on local thermodynamical equilibrium (LTE) populations, which,
however, applies only in high-density regions where collisions dominate
the excitations. Non-LTE modeling is necessary when infrared and UV
fluorescent excitation and radiative de-excitation compete with
molecular collisions \citep{vandertak:11}, and it is required for a
detailed understanding of the CO ro-vibrational lines in terms of
spatial location and efficiency of the IR/UV fluorescence
\citep{thi:13}. The non-LTE analysis requires accurate collision rate
coefficients of CO with its main collision partners, H, H$_2$, He, and
electrons, as input. The lack, or limited reliability, of such rate
coefficient data hinders non-LTE modeling of molecular spectra for many
astrophysical environments. In particular, rate coefficients for the
H-CO system are known to be highly uncertain \citep{shepler:07} and
difficult to calculate due to the existence of a chemical bond between H
and CO. The appearance of a deep well and a dissociation barrier in the
H-CO potential make it substantially more difficult to compute converged
scattering cross sections than for typical van der Waals systems such as
He-CO and H$_2$-CO. In addition, it was shown in \citet{walker:14} that
standard approaches for scaling He-CO rate coefficients to obtain values
for H$_2$-CO or, especially, H-CO are not valid.

For pure rotational transitions, rate coefficients of H-CO were first
given by \citet{chu:75} about four decades ago for CO excitation from
$j=0-4$ to $j'=1-5$ at temperatures of $T = 5-150$~K. One year later,
\citet{green:76} reported their values for transitions from initial
states with $j=0-7$ at temperatures of $T=5-100$~K. Rate coefficients
for pure rotational transitions up to $j=7$ for a broad range of gas
temperatures $T=5-3000$~K, were calculated by \citet{balakrishnan:02}.
Compared with the earlier values of \citet{green:76}, their results
differed by a factor of 30 for rate coefficients at temperatures below
100~K. More recently, \citet{shepler:07} performed scattering
calculations with two new \textit{ab initio} potential energy surfaces
and suggested that the pure rotational rate coefficients obtained by
\citet{balakrishnan:02} were incorrect, due to the fact that the WKS
interaction potential \citep{keller:96} they used for H-CO was
inaccurate in its long range part. In a recent paper, \citet{yang:13}
calculated the rotational quenching rate coefficients of low-lying
($j=1-5$) rotational CO levels for temperatures ranging from 1 to
3000~K. They found pure rotational quenching rate coefficients similar
to the values of \citet{green:76}, which confirms the inaccuracy of
results obtained using the WKS potential \citep{balakrishnan:02}. Very
recently, \citet{walker:15} performed calculations of the pure
rotational quenching rates for initial states up to $j=45$ at a
temperature range of $T =1-3000$~K based on the new three-dimensional
(3D) H-CO potential of \citet{song:13}.

For vibrational rate coefficients the situation is still unsatisfactory,
however. Experimental values are available only for the rotationally
unresolved $v=1 \rightarrow 0$ transition. The most reliable results
were measured by \citet{glass:82} using a discharge-flow shock tube at
temperatures between 840 and 2680~K. Other experimental values reported
by \citet{vonrosenberg:74} and \citet{kozlov:00} are based on estimates
of the efficiency of H atoms in vibrational relaxation of CO derived
from data involving other gases. Quantum scattering calculations of rate
coefficients for transitions between vibrational levels up to $v=4$ were
performed by \citet{balakrishnan:02} with the infinite order sudden
(IOS) approximation at temperatures ranging from 100 to 3000~K. Their
computed results agreed with the experimental data measured by
\citet{glass:82} in the high-temperature range. Vibrational rate
coefficients below 100~K were not available, however. Rate coefficients
for ro-vibrational transitions were even less complete. \citet{yang:05}
presented quenching rate coefficients for initial states $v=1, j=0,1,2$
to $v=0$ at a temperature range of $T = 10^{-5}-300$ K. They performed
close-coupling (CC) calculations, but their channel basis may have been
too small to ensure convergence and they adopted the WKS potential
\citep{keller:96}, known to be inaccurate in the long range. Both
factors may lead to uncertainties in the ro-vibrational rate
coefficients and the lack of a comprehensive set of H-CO ro-vibrational
rate coefficients obliged astronomers to use scaling laws to extrapolate
the data \citep[e.g.,][]{thi:13}.

Recently, \citet{thi:13} modeled CO ro-vibrational emission from Herbig
Ae discs using the H-CO rate coefficients calculated by
\citet{balakrishnan:02}. They extrapolated the pure rotational
de-excitation rate coefficients to initial states with $j>7$ and
vibrational transition rate coefficients to temperatures below 100~K.
Assuming complete decoupling of rotational and vibrational motion, they
estimated the state-to-state ro-vibrational rate coefficients from the
corresponding rates for pure rotational transitions in the ground
vibrational level. This method was described in detail and applied to
the H$_2$O-H$_2$ system by \citet{faure:08}. However, extrapolations
based on only a few calculated rate coefficients may cause large errors
in the deduced rates. Moreover, the above mentioned inaccuracy of the
pure rotational rate coefficients derived from the WKS potential
\citep{balakrishnan:02} will introduce an additional error. In the
present paper, we report explicitly calculated state-to-state
ro-vibrational de-excitation rate coefficients for $\Delta v=-1$
transitions in H-CO for initial states with $v=1$ and 2 and $j$ values
up to 30 for temperatures ranging from 10 to 3000~K. A new extrapolation
method is devised and applied to obtain state-to-state ro-vibrational
de-excitation rate coefficients for initial states with $v$ up to 5. A
LAMDA-type file \citep{schoier:05} is provided with ro-vibrational
de-excitation rate coefficients of H-CO for $v=1-5,j=0-30 \rightarrow
v',j'$ transitions at temperatures of $T=10-3000$ K.

\section{Theory}
\label{sec:meth}
\subsection{Equations for rate coefficients}
The CC calculations were performed using our scattering code described
by \citet{song:15}, while the IOS calculations were carried out with the
non-reactive scattering program MOLSCAT \citep{molscat:94}. The
scattering methods used to obtain cross sections are reported in detail
in our previous paper \citep{song:15}. Rate coefficients for specific
ro-vibrational transitions were calculated by averaging the
corresponding cross sections over a Maxwell-Boltzmann distribution of
translational energies of the colliding particles,
\begin{equation}
\label{eq:rate}
r(T) = \left( \frac{8k_{B}T}{\pi \mu} \right)
^{\frac{1}{2}} \frac{1}{\left( k_{B}T \right)^2} \int_{0}^{\infty}
\sigma(E_{k}) \mathrm{exp}\left(-\frac{E_{k}}{k_{B}T}\right) E_{k}
dE_{k},
\end{equation}
where $\mu$ is the reduced mass of H-CO, $k_{B}$ is the Boltzmann
constant and $T$ is the gas temperature. Cross sections were calculated
over an energy range from 0.1 to 15000 cm$^{-1}$, see
\citet{song:15}. The integral over collision energies was computed
numerically with the trapezoidal rule after cubic spline
interpolation of the cross sections on a logarithmic energy scale.
The cross section $ \sigma(E_k) $ as a function of the collision energy
$E_k$ can be the state-to-state ro-vibrational cross section $
\sigma_{v,j \rightarrow v',j'} (E_k) $, the total vibrational quenching
cross section $ \sigma_{v,j \rightarrow v'} (E_k) $ for the transitions
from an initial $v,j$ state to a final $v'$ state, or the vibrational
transition cross section $\sigma_{v \rightarrow v'} (E_k) $. The quantum
numbers $v$ and $j$ refer to the vibration and rotation in the initial
state, while $v'$ and $j'$ refer to the final state. The rate
coefficients for the reverse transitions can be obtained by detailed
balance
\begin{equation}
\label{eq:detail}
r_{v',j' \rightarrow v,j}(T) = \frac{2j+1}{2j' +1}
\mathrm{exp}\left(\frac{\epsilon_{v',j' } -
\epsilon_{v,j}}{k_{B}T}\right) r_{v,j \rightarrow v',j' }(T),
\end{equation}
where $ \epsilon_{v,j} $ and $ \epsilon_{v',j'} $ are the energies of the
ro-vibrational levels. The state-to-state ro-vibrational cross sections
$ \sigma_{v,j \rightarrow v',j'} ^{\mathrm{CC}} (E_k) $ are obtained from
full CC calculations; the corresponding rate coefficients can be
calculated from Eq.~(\ref{eq:rate}). By summation of the state-to-state
ro-vibrational cross sections over all final $j'$ levels in the
$v'$ state, we obtain the total vibrational quenching cross section from
a specific ro-vibrational initial state $v,j$ to final state $v'$
\begin{equation}
\label{eq:csvjv}
\sigma_{v,j\rightarrow v'} ^{\mathrm{CC}} (E_{k}) = \sum_{j'}
{\sigma_{v,j\rightarrow v',j'} ^{\mathrm{CC}} (E_{k})}.
\end{equation}
The corresponding rate coefficients are denoted by $ r_{v,j \rightarrow
v'} ^{\mathrm{CC}} (T) $. Averaging the rate coefficients $r_{v,j
\rightarrow v'} ^{\mathrm{CC}} (T) $ over a thermal population of
initial $j$ states yields the vibrational transition rate coefficient
based on the CC approach
\begin{equation}
\label{eq:rvv}
r_{v \rightarrow v'} ^{\mathrm{CC}} (T) = \frac{\sum_{j}{g_j \mathrm{exp}
(-\frac{\epsilon_{v,j}}{k_{B}T})r_{v,j \rightarrow v'} ^{\mathrm{CC}} (T)}}
{\sum_{j} {g_j \mathrm{exp} (-\frac{\epsilon_{v,j}}{k_{B}T})}},
\end{equation}
where $g_{j}= 2j+1$ is the degeneracy of the rotational level $j$.

The vibrational transition cross sections $\sigma_{v\rightarrow v'}
^{\mathrm{IOS}} (E_k) $ are obtained directly from scattering
calculations in the IOS approximation. The corresponding rate
coefficients can again be calculated from Eq.~(\ref{eq:rate}).

\subsection{Extrapolation method}
\label{extra}
If we wish to calculate all ro-vibrational rate coefficients of interest
for astrophysical modeling, the full CC method is prohibitively
expensive. The incompleteness of the available ro-vibrational rate
coefficients forced astronomers to use extrapolated data, commonly based
on a complete decoupling of vibration and rotation
\citep{faure:08, thi:13, bruderer:15}. Here, we introduce a new
extrapolation method for state-to-state ro-vibrational rate coefficients
in which we assume the coupling in $v,j \rightarrow v',j'$ transitions
with $v'<v$ to be the same as in the $v=1,j \rightarrow v'=0, j'$ transition. Then,
the state-to-state ro-vibrational rates are related to the corresponding
rates for the $v=1,j \rightarrow v' = 0,j'$ transitions as follows
\begin{equation}
\label{eq:extrap}
r_{v,j \rightarrow v',j'} (T) = P_{v v'} (T)
r_{1,j \rightarrow 0,j'} (T),
\end{equation}
where the factor $P_{v v'} (T)$ is defined as
\begin{equation}
\label{eq:pvv}
P_{v v'} (T) = \frac{r_{v \rightarrow v'} (T)
\sum_{j}{g_j \mathrm{exp}(-\frac{\epsilon_{v,j}}{k_{B}T})}}
{\sum_{j}{[ g_j \mathrm{exp} (-\frac{\epsilon_{v,j}}{k_{B}T})
\sum_{j'}{ r_{1,j \rightarrow 0,j'} (T) } ]}} =
  \frac{r_{v\rightarrow v'}(T)}{r_{1 \rightarrow 0} (T) }.
\end{equation}
The difference with the extrapolation method used by \citet{thi:13} and
\citet{faure:08} lies in the replacement of $r_{0,j \rightarrow 0,j' }$
by $ r_{1,j \rightarrow 0,j'} $. Although this change seems minor, the
improvement in the estimated rate coefficients is substantial. We will show
this in detail in Sec.~\ref{sec:resu}, where we compare the extrapolated
rates of $v=2,j \rightarrow v'=1,j'$ transitions with results
obtained directly from CC calculations. The ro-vibrational rate
coefficients $ r_{1,j \rightarrow 0,j'}(T) $ have been calculated using
the full CC method, while the vibrational rate coefficients $ r_{v
\rightarrow v'} (T) $ were obtained from the IOS approximation. All
other ro-vibrational rate coefficients $ r_{v,j \rightarrow v',j'} (T) $
can then be obtained from the extrapolation formula, Eq.~(\ref{eq:extrap}).
Another advantage of our method is that it also yields the values for
$v,j \rightarrow v' ,j'=j$ transitions; these could not be obtained with
the previous extrapolation method since the data for rotationally
elastic $j \rightarrow j$ transitions were not tabulated. And even if
they had been available, they most likely would have given
much too large extrapolated results.

\section{Results}
\label{sec:resu}
\subsection{Rate coefficients from quantum scattering calculations}
\begin{figure}
\epsscale{.90}
\plotone{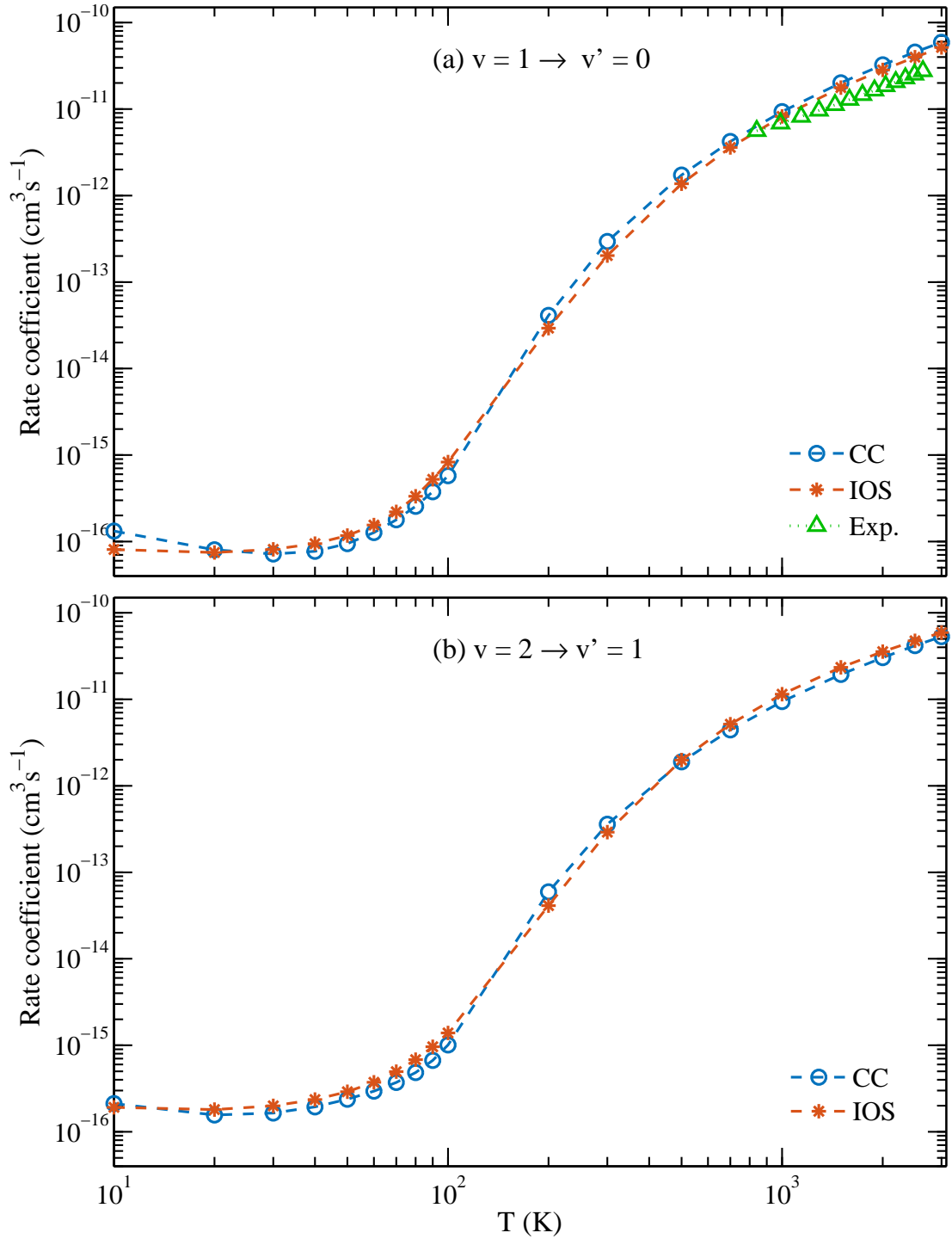}
\caption{\label{fig:vv}
Comparison of CC and IOS rates
for vibrational quenching $v = 1 \rightarrow v' = 0$ and
$v = 2 \rightarrow v' = 1$. The experimental rates are from
\citet{glass:82}.}
\end{figure}

Cross sections for ro-vibrational $v=1,j=0-30 \rightarrow v'=0,j'$
transitions from full CC calculations are reported in our previous paper
\citep{song:15}. We also concluded in that paper that the coupled states
approximation is only suitable for collision energies above
$\approx$1000 cm$^{-1}$ for the H-CO system and it is therefore not
applied here. Table~\ref{tab:rovir} lists the ro-vibrational transition
rate coefficients for a gas temperature range of 10~K $ \leq T \leq $
3000~K calculated from these cross sections with the use of
Eq.~(\ref{eq:rate}). The highest final $j'= 27- 42$ values for which the
rates are given in this table depend on the initial $j$ quantum number.
Transitions for even larger final $j'$ are not reported, either because
they are negligibly small, or because they were not completely converged
with the largest number of partial waves, i.e., the highest total
angular momentum $J$ included. However, we add all final $j'$ states,
including those that may not be fully converged, to calculate the
vibrational quenching cross sections $\sigma_{v,j \rightarrow v'}
^{\mathrm{CC}} (E_k)$ and the vibrational transition rate coefficients
$r_{v \rightarrow v' } ^{\mathrm{CC}}(T)$. Hence, the CC vibrational
quenching rate coefficients shown in Fig.~\ref{fig:vv} are slightly
different from the values that would be obtained by summing and
averaging only the reported rates $r_{v,j \rightarrow v',j'}
^{\mathrm{CC}}(T)$.

In order to test our proposed extrapolation scheme, we also calculated
cross sections and rate coefficients for ro-vibrational $v=2,j=0-30
\rightarrow v' =1,j'$ transitions with the CC method. We used a channel
basis $B4(75,65,54,41,20)$, where the notation $Bn(j_0,j_1,j_2,...,j_n)$
represents a basis with the highest vibrational level $n$ and the
highest rotational level $j_i$ for vibrational level $v=i$
\citep[see][]{song:15}. The state-to-state ro-vibrational rate
coefficients were computed for a temperature range of 10~K $ \leq T \leq
$ 3000~K; they are listed in Table~\ref{tab:rovir2}.

In addition, we obtained vibrational quenching cross sections from
scattering calculations in the IOS approximation with a $v=0-14$ basis
for CO. This basis of 15 vibrational levels, with energies given in
Table \ref{tab:viblev}, is sufficiently large to converge all $v=1-5
\rightarrow v' < v$ transitions. In Table~\ref{tab:vibr} we report the
IOS vibrational transition rate coefficients for a temperature range of
10~K $ \leq T \leq $ 3000~K. The IOS approximation is not always
suitable for cross sections at low collision energy and for rate
coefficients at low temperature. We check this by comparing the
vibrational rate coefficients produced with the IOS approximation and
the results obtained by summing the state-to-state CC
rate coefficients over final rotational $j'$ levels and averaging over
initial $j$ levels. Figure~\ref{fig:vv} shows that the IOS rate
coefficients agree well with the CC results, for both $v=1 \rightarrow
v'=0$ and $v=2 \rightarrow v'=1$ transitions. The calculated results for
the $v=1 \rightarrow v'=0$ transition also agree with the experimental
data measured for this transition by \citet{glass:82}. All deviations
between the IOS and CC rate coefficients in Fig.~\ref{fig:vv} are less
than 45\%, which is sufficiently accurate for astrophysical
applications. The IOS rate coefficients for $v=4 \rightarrow v' = 1,2,3$
transitions in Table~\ref{tab:vibr} at temperatures $T=40$ and 50~K were
obtained by interpolation of the rate coefficients at other
temperatures. The original IOS data show a sharp peak around these
temperatures due to resonance effects in the cross sections.
However, the resonances in the IOS cross sections do not precisely
coincide with the resonances in the cross sections from CC calculations
\citep[see Fig.~2 of][]{song:15}.

\subsection{Extrapolated rate coefficients}

\begin{figure}
\epsscale{1.0}
\plotone{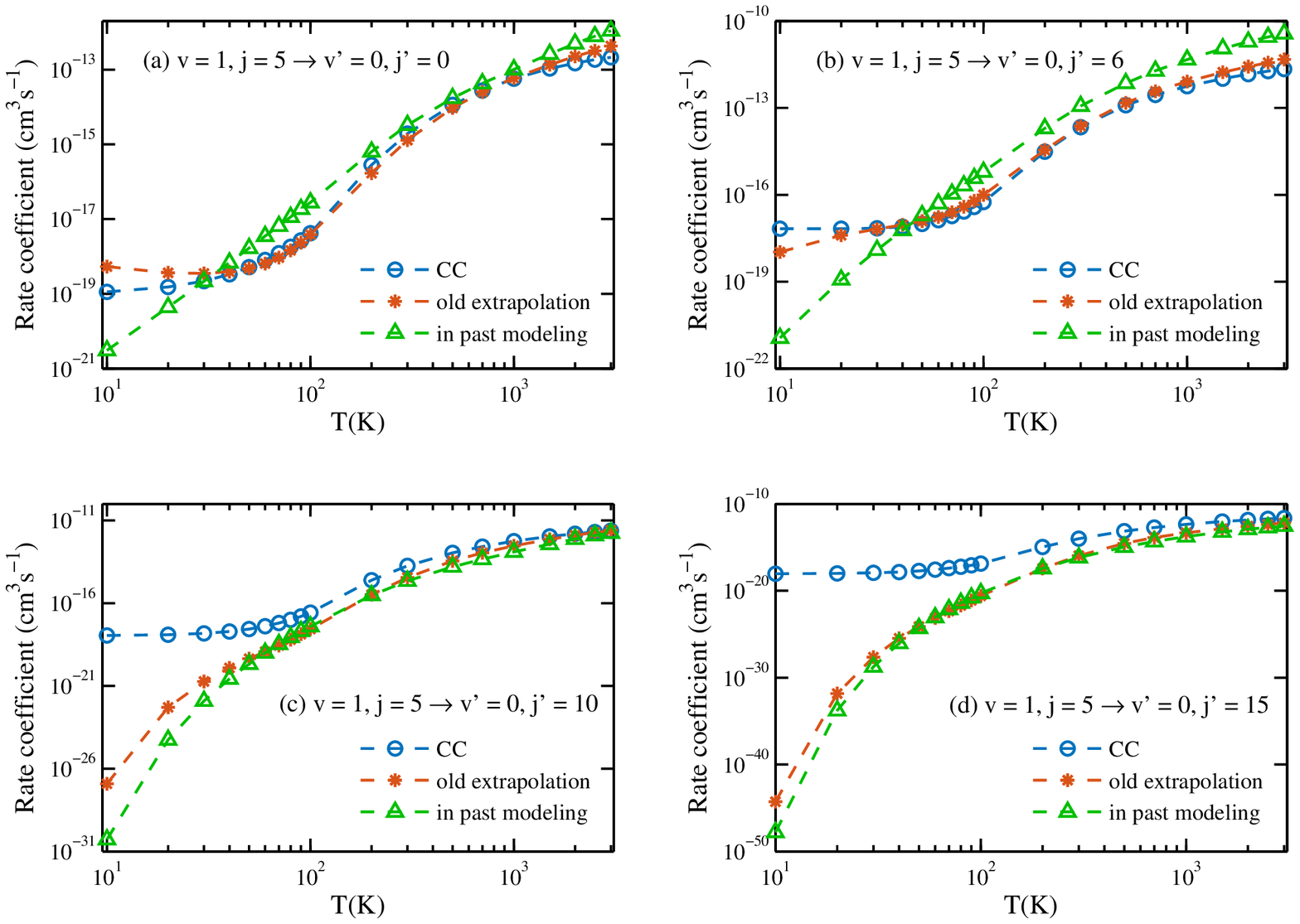}
\caption{\label{fig:ccespure}
Comparison of ro-vibrational transition rate coefficients from CC
calculations with extrapolated data obtained with the old extrapolation
method of \citet{thi:13} and \citet{faure:08} (using vibrationally
elastic data) and with the data used in astrophysical modeling \citep{thi:13}.}
\end{figure}

First, we test the validity of the extrapolation method used by
\citet{thi:13} and \citet{faure:08}. Combining the IOS vibrational rate
coefficient for the $v=1 \rightarrow v' =0$ transition with pure
rotational (vibrationally elastic) rate coefficients from
\citet{walker:15}, both of them based on the 3D H-CO potential of
\citet{song:13}, we can obtain the state-to-state ro-vibrational rate
coefficients for $v=1,j=0-30 \rightarrow v' =0, j'$ transitions by
extrapolation. Examples of the comparison of our CC rate coefficients,
rate coefficients from extrapolation based on the method of
\citet{faure:08} and \citet{thi:13}, and the corresponding rate
coefficients used by \citet{thi:13} are illustrated in
Fig.~\ref{fig:ccespure}. For transitions with final $j'$ less than or
close to the initial $j$, i.e., the $v=1, j= 5 \rightarrow v' =0,j'
=0,6$ transitions, the CC and extrapolated rate coefficients agree quite
well with each other, as shown in Figs.~\ref{fig:ccespure}(a) and (b).
Large discrepancies appear, however, between CC and extrapolated rate
coefficients for transitions with larger $\Delta j = j' -j$, especially
at lower temperatures, see Figs.~\ref{fig:ccespure}(c) and (d). These
large discrepancies result because the pure rotational (vibrationally
elastic) transitions in $v=0$ are endoergic for $j' > j$, while the
ro-vibrational quenching processes $v=1,j \rightarrow v'=0,j'$ are
exoergic for all $j'<33$ even when $j=0$. In addition, by analogy with
Eq.~(\ref{eq:pvv}), the vibration-related scale factor $P_{vv'}(T) (v
\neq v')$ can be rewritten in the extrapolation method of \citet{thi:13}
and \citet{faure:08} as $r_{v\rightarrow v'}(T)/r_{0\rightarrow 0}(T)$.
Clearly, the validity of scaling inelastic vibrational transition rate
coefficients with vibrationally elastic data is questionable. The rate
coefficients adopted in the modeling of \citet{thi:13} are even more
discrepant, as shown by the green lines with triangle markers in
Fig.~\ref{fig:ccespure}. In their extrapolations, the pure rotational
rates were adopted from \citet{balakrishnan:02} who used the WKS
potential, known to be inaccurate at long range. Moreover, the pure
rotational rates for initial states with $j > 7$ and the vibrational
rates for temperatures below 100~K were not available and had to be
obtained by extrapolation.

\begin{figure}
\epsscale{1.0}
\plotone{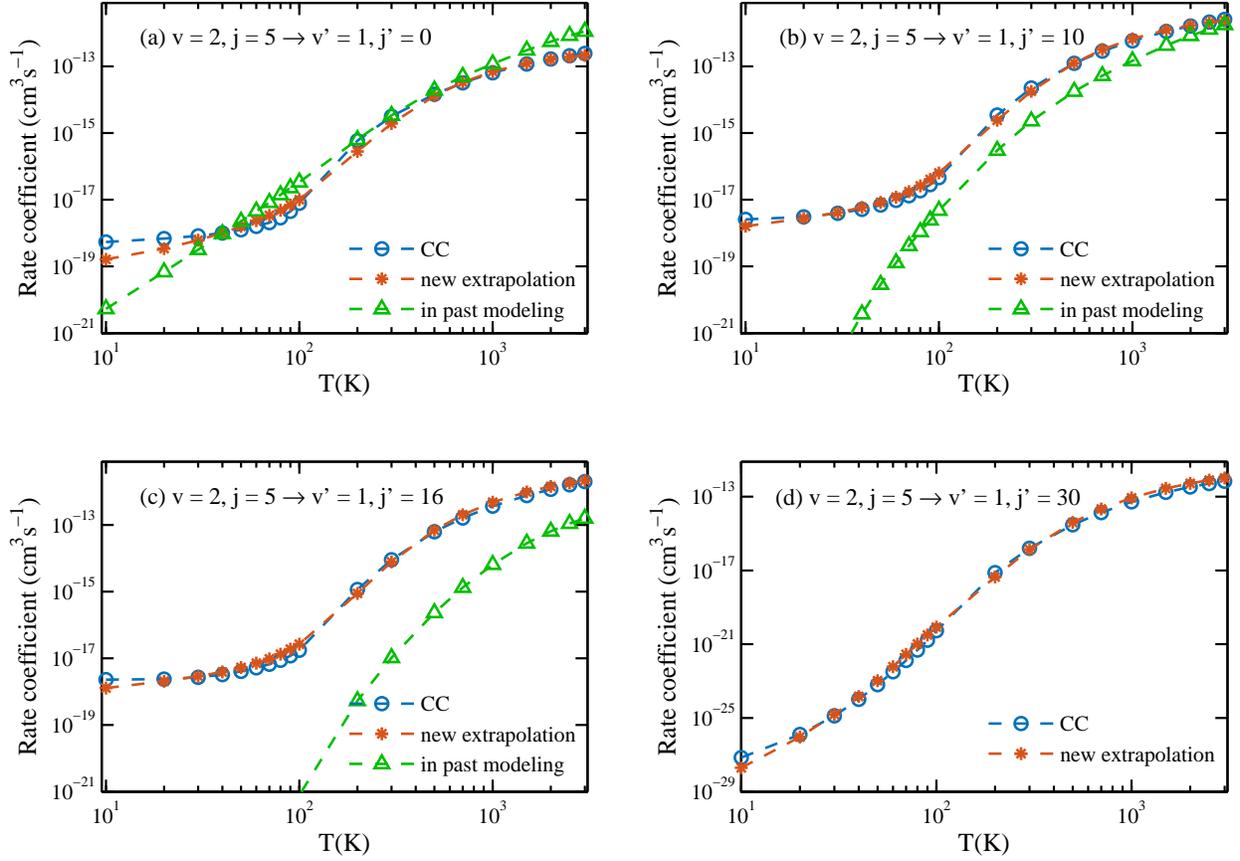}
\caption{\label{fig:ccesas}
Comparison of ro-vibrational transition rate coefficients from CC
calculations with extrapolated data obtained with our new extrapolation
method (using vibrationally inelastic data) and with the data used in
astrophysical modeling \citep{thi:13}.}
\end{figure}

Let us now discuss the results from our new extrapolation method based
on vibrationally inelastic rates from quantum scattering calculations.
Using the data in Tables~\ref{tab:rovir} and \ref{tab:vibr}, the
ro-vibrational rate coefficients for $v=2-5,j=0-30 \rightarrow v' < v,
j'$ transitions can be extrapolated with Eq.~(\ref{eq:extrap}).
Figure~\ref{fig:ccesas} illustrates some comparisons of ro-vibrational
rate coefficients for transitions from $v=2, j$ to $v' = 1, j'$. It
shows the full CC results, the rate coefficients from the new
extrapolation method based on vibrationally inelastic data, and the
results of \citet{thi:13} based on vibrationally elastic data. For the
four transitions shown, the new extrapolated rate coefficients agree
very well with those from full CC calculations. The root mean square
relative deviation of the rate coefficients in the temperature range of
10~K $\leq T \leq$ 3000~K for all the transitions $v=2,j=0-30
\rightarrow v' =1,j'$ is 40\%. This is sufficiently accurate for
astronomical applications. However, the data from the scaling
approach of \citet{thi:13} deviate
dramatically from our results, especially for higher final $j'$ values.
The reasons why their data are less accurate were already discussed
above. The deviation is largest in the temperature range of 10~K $ \leq
T \leq $ 100~K, probably because the vibrational transition rate
coefficients at low temperature were not available at the time.

\section{Astrophysical models}
In order to show the relevance of these new collision rate coefficients in
the astrophysical context, we chose a twofold approach: (1) simple 1D
slab models with constant temperature, density and CO abundance and (2)
the 2D radiative thermo-chemical disk model from \citet{thi:13}. In the
following, we compare the results obtained with the old H-CO
collision rates to those obtained with the new data. Since this paper
presents new collision rates for $v \leq 5$ and $j \leq 30$, we
restricted the calculations below to the same range of quantum numbers.

\subsection{ProDiMo slab and disk models}

ProDiMo is a 2D radiation thermo-chemical disk code \citep{woitke:09}.
The code solves the 2D continuum radiative transfer to obtain the dust
temperature and radiation field throughout the disk and provides the
self-consistent solution for the chemistry and gas heating/cooling
balance. The gas temperatures can be used iteratively to find a
vertical hydrostatic equilibrium solution. However, we use for this work
a fixed parametrized gas scale height and assume that gas and dust are
well mixed. Details of the numerical methods, the chemical network and
a list of heating/cooling processes can be found in \citet{woitke:09}.
The CO molecular data used are described in detail above and in
\citet{thi:13}.

The code can be used in 1D slab mode, to calculate the emission emerging
from a fixed total gas column with a constant gas temperature and
constant volume densities of collision partners (H, H$_2$, He and
e$^-$). The non-LTE level populations are calculated using the
escape probability method taking into account also IR pumping by thermal
dust emission. Since we focus here on the effect of the new collisional
rates, we minimize the effect of IR pumping by fixing the dust
temperature at 20~K. The turbulent width is fixed to $v_{\rm
turb}=1.0$~km/s and the total broadening is given by $b = \sqrt{v_{\rm
th}^2 + v_{\rm turb}^2}$, where $v_{\rm th}$ is the thermal velocity of
the CO molecules. The slab models provide a powerful means to exploit a
very wide parameter space under which CO emission could arise in space.

In addition, we use the standard model for a disk around a
$2.2$~M$_\odot$ Herbig star ($L_\ast = 32$~L$_\odot$) taken from
\citet{thi:13}. The disk has a radial size of 300~AU and a mass of
$10^{-2}$~M$_\odot$. More details on the dust opacities and vertical
disk structure can be found in Table~3 of \citet{thi:13}. Such a disk
model does not provide any freedom as to the conditions under which CO
is emitting. Its abundance, excitation and emission are calculated
self-consistently given the density distribution of gas within the disk and
the irradiating stellar and interstellar radiation field.

\subsection{Slab model results}

We ran four series of models, each with three different total
gas column densities of $N_{\rm \langle H
\rangle}\!=\!10^{19}$, $10^{21}$, and $10^{23}$~cm$^{-2}$.
The CO abundance is fixed at $10^{-4}$ with respect to the total
hydrogen number density. Series 1 uses a fixed gas temperature $T$ of
800~K and high densities of collision partners, $\log n_{\rm
H}\!=\!\log n_{\rm H_2}\!=\!9$, $\log n_{\rm He}\!=\!8$, $\log n_{\rm
e}\!=\!5$ (in cm$^{-3}$). Note that the gas in these slab models is
not fully molecular, i.e., the H/H$_2$ abundance ratio is 1. Series
2 uses a lower gas temperature of $200$~K and the same collision partner
densities. Series 3 uses $T\!=\!800$~K and a lower density of collision
partners, $\log n_{\rm H}\!=\! \log n_{\rm H_2}\!=\!6$, $\log
n_{\rm He}\!=\!5$, $\log n_{\rm e}\!=\!2$ (in cm$^{-3}$).
An additional series (4) was calculated to isolate the effect of
H-collisions. In this series, the temperature and density of H are
the same as in series 3, while the collision partner densities of
H$_2$, He and electrons are assumed to be negligible. 

\begin{figure}
\plotone{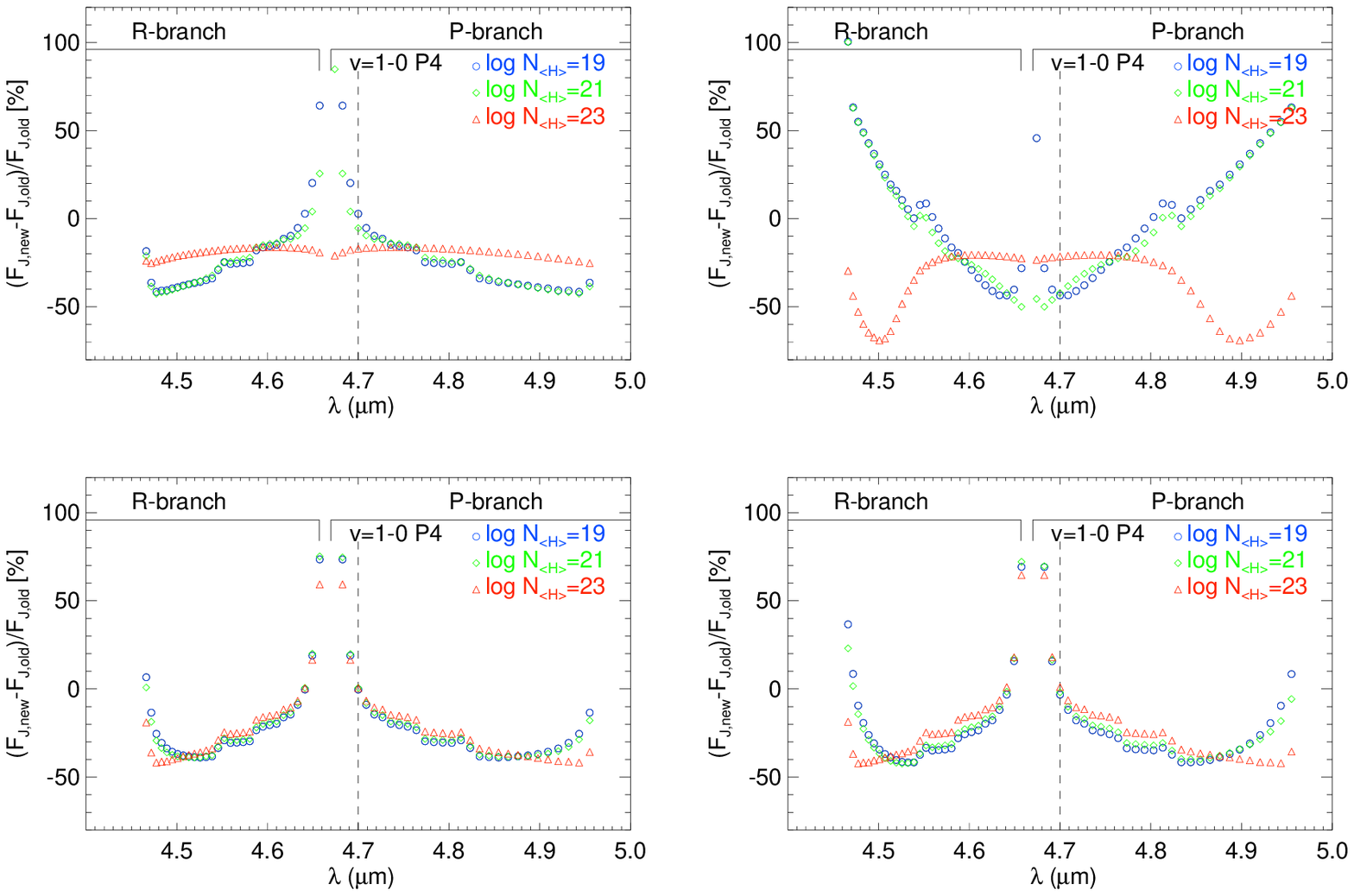}
\caption{\label{fig:COslab}
Changes in the modeled CO ro-vibrational emission from a series of slab
models using the new and previously available H-CO rate coefficients.
Top left (series 1): $T\!=\!800$~K, $\log n_{\rm H,H_2}\!=\!9$, $\log
n_{\rm He}\!=\!8$, $\log n_{\rm e}\!=\!5$ in units of cm$^{-3}$.
Top right (series 2): $T\!=\!200$~K, $\log n_{\rm H,H_2}\!=\!9$,
$\log n_{\rm He}\!=\!8$, $\log n_{\rm e}\!=\!5$ - note the
different vertical scale. Bottom left (series 3): $T\!=\!800$~K, $\log
n_{\rm H,H_2}\!=\!6$, $\log n_{\rm He}\!=\!5$, $\log n_{\rm
e}\!=\!2$. Bottom right (series 4): $T\!=\!800$~K, $\log n_{\rm
H}\!=\!6$, collisions with other partners negligible. The total gas
column density of the slab $N_{\rm \langle H \rangle}$ is given in the
legend.}
\end{figure}

Figure~\ref{fig:COslab} illustrates the change in the non-LTE fluxes of
the $v=1-0$ band resulting from the old and new collisional rates for
H-CO in the slab models. We note that changes are typically smallest for
the slab with the largest line optical depth $\log N_{\rm \langle H
\rangle}\!=\!23$ ($\log N_{\rm CO}\!=\!19$). For slabs at $T\!=\!800$~K,
the line fluxes with the new rates are smaller than those with the old
rates. Exceptions are the smaller and the highest $j$ levels; the flux
changes are generally largest for the lowest four levels. It is also
evident from these comparisons that the implementation of the new rates
will have an effect on the shape of the vibrational band structure.
In most cases of high $T$ (800 K) slabs, the band structure (flux
vs wavelength) will get flatter for low $j$ and drop faster at high $j$
compared to the band structure with the old collision data. The total
cooling rate of the $v=1-0$ band ($j\!\leq\!30$) is typically 20-30\%
lower with the new collisional rates. This will have an impact on
the gas temperature whenever CO cooling dominates the energy balance.
Examples are (1) the surface layers of the inner $\approx\,10$~pc of AGN
disks \citep{meijerink:13}, (2) the inner $\approx\,10$~AU of
protoplanetary disks \citep{woitke:09} and (3) intermediate density
($n_{\langle \rm H \rangle}\ \approx\,10^5$~cm$^{-3}$) J-type shocks
\citep{hollenbach:89}. If H collisions are omitted entirely in slab
series 1, the fluxes in the most optically thick case are lower by a
factor of 100; hence H collisions are more important than H$_2$
collisions and they are key in modeling the CO ro-vibrational emission. 

The gas temperature $T$ is a free parameter and we can explore the
effect of the new collision rates at lower temperatures. The comparison
between series 1 and 2 (top left and right panel of
Fig.~\ref{fig:COslab}) clearly shows that the changes in the line fluxes
increase substantially for $T\!=\!200$~K. However, at these low
temperatures, the populations in the CO vibrational levels are very low
and hence also the emerging line fluxes will be low. An exception may be
the case of fluorescent excitation, where the vibrational levels are
populated by an external radiation field to non-LTE values. Examples are
the inner disks of Herbig Ae stars, where the stellar radiation field
can substantially pump the vibrational levels to vibrational
temperatures of a few thousand K \citep{brittain:09,vanderPlas2015}.

The bottom panels of Fig.~\ref{fig:COslab} illustrate the effects for
low density environments (non-LTE cases). Here, line-to-line differences
are larger than in the high density environment. The overall picture
does not change very much if we neglect the other collision partners,
H$_2$, He, and electrons.

Besides the application to protoplanetary disks discussed in the next section,
these new rate coefficients will also be important for proper modeling of CO infrared
emission from other interstellar regions with large H/H$_2$ transition
zones such as dissociative J-type shocks \citep{neufeld:89, gonzalez:02}
or dense X-ray irradiated gas found near Active Galactic Nuclei.

\subsection{Disk model results}
\begin{figure}
\epsscale{0.8}
\plotone{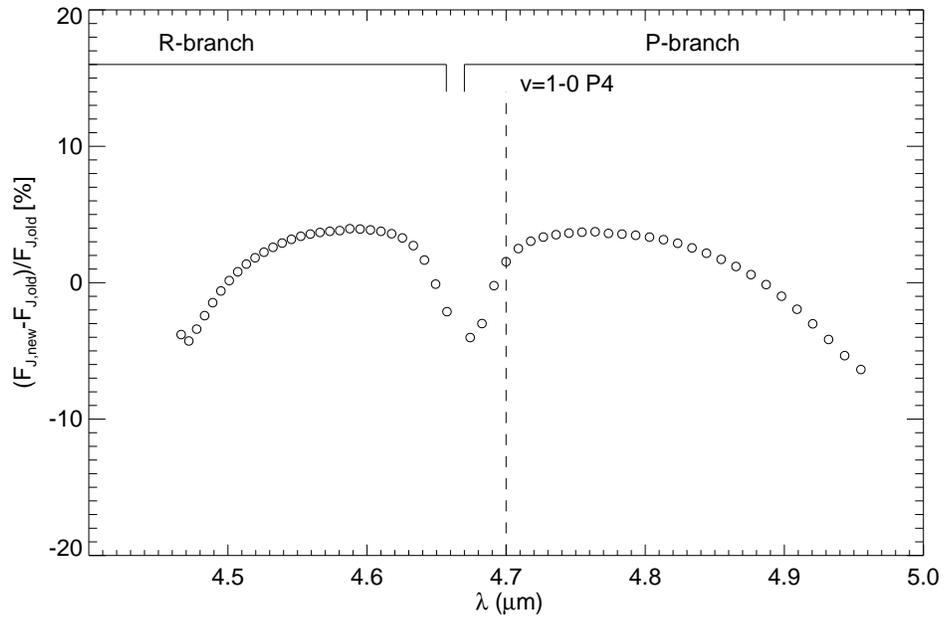}
\caption{\label{fig:diskcompare}
Changes in the modeled CO ro-vibrational emission from the disk around a Herbig Ae star
using the new and previously available H-CO rate coefficients.}
\end{figure}

Figure~\ref{fig:diskcompare} shows the change in the CO ro-vibrational
emission of the $v=1\!-\!0$ band at $4.7~\mu$m resulting from the old
and the new (this work) H-CO collision rate coefficients. The collision
rate coefficients with H$_2$, He, and electrons remain unchanged.
Differences are small, on the order of $\pm 5$\%. The line fluxes with
the new rates show a clear pattern with respect to the old ones: the new
rates lead to lower line fluxes at low $j$, higher fluxes in the
intermediate range and then again lower fluxes for $j\!\gtrsim\!25$. The
upturn at the highest $j$ is a boundary effect, since the rotational
states of CO are artificially cut off in the model at $j\!=\!30$. As
noted in \citet{thi:13}, the CO emission arises from a hot surface layer
with $T\approx\!1000$~K, where the gas is predominantly atomic (see
their Fig.12). At these high temperatures, differences between the old
and new collision rates are smaller than at low temperatures
(Fig.~\ref{fig:ccespure}) and this partially explains the
relatively small changes observed in the disk model. In addition, IR
pumping is competing with collisions, especially in the inner disk,
where dust continuum emission peaks in the near-IR ($1000~{\rm
K}\!<\!T_{\rm dust}\!<\!1500$~K).

\section{Conclusions}
\label{sec:concl}
We present ro-vibrational de-excitation rate coefficients from full CC
quantum scattering calculations for $v=1,j=0-30 \rightarrow v' =0,j'$
and $v=2,j=0-30 \rightarrow v'=1, j'$ transitions of CO in collision
with H atoms in the temperature range of 10~K $ \leq T \leq $ 3000~K.
Also vibrational quenching rate coefficients from IOS calculations are
given for $v=1-5 \rightarrow v' < v$ transitions in the same temperature
range. We propose a new extrapolation method to obtain the
ro-vibrational rate coefficients for $v=2-5,j=0-30 \rightarrow v'<v,j'$
transitions from our quantum mechanically calculated results. Comparison
of the CC ro-vibrational rate coefficients for the $v=2, j=0-30
\rightarrow v'=1, j'$ transitions with the extrapolated ones confirms
the reliability of this extrapolation method. A LAMDA-type
\citep{schoier:05} file with state-to-state ro-vibrational rate
coefficients for $v=1-5,j=0-30 \rightarrow v' < v,j'$ transitions is
provided online. Compared with the ro-vibrational rate coefficients for
H-CO used previously in astrophysical modeling \citep{thi:13}, the
current dataset is both comprehensive and accurate. Application of the
previously available and the new data set in slab models shows that
typical changes in line fluxes under astrophysical conditions are on the
order of 20-70\% in high temperature environments (800~K) with an
H/H$_2$ ratio of 1 and increase for lower temperatures. In the inner
disks around Herbig Ae stars, the line flux changes are smaller due to
the very high temperatures and densities under which this emission
arises.

\acknowledgments
We are grateful to Ewine van Dishoeck for suggesting this study and we
thank her, Simon Bruderer, and Fran\c{c}ois Lique for valuable
discussions. We thank the computer and communication department (C\&CZ)
of the Faculty of Science of the Radboud University for computer
resources and technical support. The work is supported by The
Netherlands Organisation for Scientific Research, NWO, through the Dutch
Astrochemistry Network, and in part by the National Science Foundation
under Grant No. NSF PHY11-25915. N.~B. is supported in part by NSF grant
PHY-1505557. K.M.~W. and P.C.~S. acknowledge support from NASA grants
NNX12AF42G and NNX13AF42G. I.~K. and W.F.~T. acknowledge funding from
the European Union Seventh Framework Programme FP7-2011 under grant
agreement no. 284405.


\clearpage

\begin{landscape}
\begin{table}
\begin{tiny}
\caption{Rate coefficients from CC calculations for ro-vibrational
$ v=1,j=0-30 \rightarrow v'=0,j'$ transitions in the temperature
range of 10~K $ \leq T \leq $ 3000~K (in units of cm$^3$s$^{-1}$)}
\label{tab:rovir}
\begin{center}
\setlength{\tabcolsep}{1pt}
\begin{tabular}{ccccccccccccccccccccc}
\tableline\tableline
\multicolumn{2}{c}{}
&   \multicolumn{19}{c}{T (K)} \\
\cline{3-21} \\
\noalign{\vspace{-0.3cm}}
\multicolumn{1}{c}{$ j $}
&  \multicolumn{1}{c}{$ j^\prime $}
&  \multicolumn{1}{c}{ 10 }
&  \multicolumn{1}{c}{ 20 }
&  \multicolumn{1}{c}{ 30 }
&  \multicolumn{1}{c}{ 40 }
&  \multicolumn{1}{c}{ 50 }
&  \multicolumn{1}{c}{ 60 }
&  \multicolumn{1}{c}{ 70 }
&  \multicolumn{1}{c}{ 80 }
&  \multicolumn{1}{c}{ 90 }
&  \multicolumn{1}{c}{ 100 }
&  \multicolumn{1}{c}{ 200 }
&  \multicolumn{1}{c}{ 300 }
&  \multicolumn{1}{c}{ 500 }
&  \multicolumn{1}{c}{ 700 }
&  \multicolumn{1}{c}{ 1000 }
&  \multicolumn{1}{c}{ 1500 }
&  \multicolumn{1}{c}{ 2000 }
&  \multicolumn{1}{c}{ 2500 }
&  \multicolumn{1}{c}{ 3000 }\\
\tableline
0 & 0 & 1.949e-17 & 1.600e-17 & 1.585e-17 & 1.733e-17 & 2.019e-17 & 2.449e-17 & 3.032e-17 & 3.770e-17 & 4.674e-17 & 5.777e-17 &  8.020e-16 & 6.025e-15 & 4.334e-14 & 1.068e-13 & 2.096e-13 & 3.622e-13  & 4.999e-13 & 6.316e-13 & 7.573e-13 \\
0 & 1 & 2.427e-17 & 1.627e-17 & 1.346e-17 & 1.281e-17 & 1.388e-17 & 1.679e-17 & 2.172e-17 & 2.881e-17 & 3.840e-17 & 5.135e-17 & 1.440e-15 & 1.111e-14 & 7.466e-14 & 1.793e-13 & 3.488e-13 & 5.965e-13 & 8.046e-13 & 9.865e-13 & 1.148e-12 \\
0 & 2  & 2.768e-17 & 1.751e-17 & 1.338e-17 & 1.170e-17 & 1.152e-17 & 1.267e-17 & 1.514e-17 & 1.900e-17 & 2.455e-17 & 3.265e-17 &  1.297e-15 & 9.571e-15 & 5.504e-14 & 1.260e-13 & 2.465e-13 & 4.330e-13 & 5.900e-13 & 7.258e-13 & 8.462e-13  \\
\tableline
\end{tabular}
\end{center}
\end{tiny}
\tablecomments{Table~\ref{tab:rovir} is
published in its entirety in
the electronic edition of
the Astrophysical Journal. A portion is shown here
for guidance regarding its
form and content.}
\end{table}
\end{landscape}

\clearpage

\begin{landscape}
\begin{table}
\begin{tiny}
\caption{Rate coefficients from CC calculations for ro-vibrational
$ v=2,j=0-30 \rightarrow v'=1,j'$ transitions in the temperature
range of 10~K $ \leq T \leq $ 3000~K (in units of cm$^3$s$^{-1}$)}
\label{tab:rovir2}
\begin{center}
\setlength{\tabcolsep}{1pt}
\begin{tabular}{ccccccccccccccccccccc}
\tableline\tableline
\multicolumn{2}{c}{}
&   \multicolumn{19}{c}{T (K)} \\
\cline{3-21} \\
\noalign{\vspace{-0.3cm}}
\multicolumn{1}{c}{$ j $}
&  \multicolumn{1}{c}{$ j^\prime $}
&  \multicolumn{1}{c}{ 10 }
&  \multicolumn{1}{c}{ 20 }
&  \multicolumn{1}{c}{ 30 }
&  \multicolumn{1}{c}{ 40 }
&  \multicolumn{1}{c}{ 50 }
&  \multicolumn{1}{c}{ 60 }
&  \multicolumn{1}{c}{ 70 }
&  \multicolumn{1}{c}{ 80 }
&  \multicolumn{1}{c}{ 90 }
&  \multicolumn{1}{c}{ 100 }
&  \multicolumn{1}{c}{ 200 }
&  \multicolumn{1}{c}{ 300 }
&  \multicolumn{1}{c}{ 500 }
&  \multicolumn{1}{c}{ 700 }
&  \multicolumn{1}{c}{ 1000 }
&  \multicolumn{1}{c}{ 1500 }
&  \multicolumn{1}{c}{ 2000 }
&  \multicolumn{1}{c}{ 2500 }
&  \multicolumn{1}{c}{ 3000 }\\
\tableline
0 & 0 & 3.814e-17 & 3.269e-17 & 3.445e-17 & 3.888e-17 & 4.467e-17 & 5.139e-17 & 5.892e-17 & 6.746e-17 & 7.789e-17 & 9.222e-17 &  1.654e-15 & 9.767e-15 & 5.489e-14 & 1.298e-13 & 2.628e-13 & 4.838e-13 & 6.837e-13 & 8.609e-13 & 1.016e-12 \\
0 & 1 & 5.241e-17 & 3.537e-17 & 3.243e-17 & 3.475e-17 & 3.970e-17 & 4.645e-17 & 5.478e-17 & 6.516e-17 & 7.932e-17 & 1.011e-16 &  2.996e-15 & 1.754e-14 & 9.173e-14 & 2.081e-13 & 4.087e-13 & 7.353e-13 & 1.026e-12 & 1.283e-12 & 1.509e-12 \\
0 & 2 & 7.974e-17 & 4.962e-17 & 3.823e-17 & 3.441e-17 & 3.412e-17 & 3.588e-17 & 3.920e-17 & 4.433e-17 & 5.256e-17 & 6.688e-17 &  2.499e-15 & 1.466e-14 & 6.790e-14 & 1.418e-13 & 2.623e-13 & 4.562e-13 & 6.375e-13 & 8.091e-13 & 9.696e-13 \\
\tableline
\end{tabular}
\end{center}
\end{tiny}
\tablecomments{Table~\ref{tab:rovir2} is
published in its entirety in
the electronic edition of
the Astrophysical Journal. A portion is shown here
for guidance regarding its
form and content.}
\end{table}
\end{landscape}

\clearpage
\begin{table}
\caption{CO vibrational energy levels.}
\label{tab:viblev}
\begin{center}
\begin{tabular}{cccccc}
\tableline\tableline
\multicolumn{1}{c}{$ v $}
&  \multicolumn{1}{c}{$ \epsilon_{v}(\mathrm{cm}^{-1}) $}
&  \multicolumn{1}{c}{$ v $}
&  \multicolumn{1}{c}{$ \epsilon_{v}(\mathrm{cm}^{-1}) $}
&  \multicolumn{1}{c}{$ v $}
&  \multicolumn{1}{c}{$ \epsilon_{v}(\mathrm{cm}^{-1}) $} \\
\tableline
   0      &    1082   &    5    &  11534    &   10   &  21331   \\
   1      &    3225   &    6    &  13546    &   11   &  23213   \\
   2      &    5342   &    7    &  15531    &   12   &  25069   \\
   3      &    7432   &    8    &  17490    &   13   &  26899   \\
   4      &    9496   &    9    &  19424    &   14   &  28704   \\
\tableline
\end{tabular}
\end{center}
\end{table}

\clearpage


\begin{landscape}
\begin{table}
\begin{tiny}
\caption{Rate coefficients from IOS calculations for vibrational
$v=1-5 \rightarrow v' < v$ transitions in the temperature
range of 10~K $ \leq T \leq $ 3000~K (in units of cm$^3$s$^{-1}$).
Rates for transitions for $v=4 \rightarrow v'=1,2,3$
at 40 and 50~K are obtained by spline interplation.}
\label{tab:vibr}
\begin{center}
\setlength{\tabcolsep}{1pt}
\begin{tabular}{ccccccccccccccccccccc}
\tableline\tableline
\multicolumn{2}{c}{}
&   \multicolumn{19}{c}{T (K)} \\
\cline{3-21} \\
\noalign{\vspace{-0.3cm}}
\multicolumn{1}{c}{$v$}
&  \multicolumn{1}{c}{$v'$}
&  \multicolumn{1}{c}{ 10 }
&  \multicolumn{1}{c}{ 20 }
&  \multicolumn{1}{c}{ 30 }
&  \multicolumn{1}{c}{ 40 }
&  \multicolumn{1}{c}{ 50 }
&  \multicolumn{1}{c}{ 60 }
&  \multicolumn{1}{c}{ 70 }
&  \multicolumn{1}{c}{ 80 }
&  \multicolumn{1}{c}{ 90 }
&  \multicolumn{1}{c}{ 100 }
&  \multicolumn{1}{c}{ 200 }
&  \multicolumn{1}{c}{ 300 }
&  \multicolumn{1}{c}{ 500 }
&  \multicolumn{1}{c}{ 700 }
&  \multicolumn{1}{c}{ 1000 }
&  \multicolumn{1}{c}{ 1500 }
&  \multicolumn{1}{c}{ 2000 }
&  \multicolumn{1}{c}{ 2500 }
&  \multicolumn{1}{c}{ 3000 }\\
\tableline
1 & 0 & 8.076e-17 & 7.485e-17 & 8.113e-17 & 9.464e-17 & 1.172e-16 & 1.550e-16 & 2.205e-16 & 3.335e-16 & 5.233e-16 & 8.297e-16 & 2.929e-14 & 2.022e-13 & 1.372e-12 & 3.594e-12 & 8.206e-12 & 1.768e-11 & 2.833e-11 & 3.970e-11 & 5.148e-11 \\
2 & 1 & 1.914e-16 & 1.808e-16 & 1.995e-16 & 2.355e-16 & 2.906e-16 & 3.728e-16 & 4.960e-16 & 6.817e-16 & 9.628e-16 & 1.388e-15 & 4.115e-14 & 2.911e-13 & 1.981e-12 & 5.144e-12 & 1.145e-11 & 2.339e-11 & 3.554e-11 & 4.746e-11 & 5.899e-11 \\
2 & 0 & 1.164e-17 & 1.107e-17 & 1.284e-17 & 1.629e-17 & 2.189e-17 & 3.076e-17 & 4.478e-17 & 6.708e-17 & 1.028e-16 & 1.601e-16 & 7.337e-15 & 5.767e-14 & 4.257e-13 & 1.175e-12 & 2.858e-12 & 6.689e-12 & 1.141e-11 & 1.674e-11 & 2.243e-11 \\
3 & 2 & 3.312e-16 & 3.008e-16 & 3.201e-16 & 3.648e-16 & 4.352e-16 & 5.387e-16 & 6.902e-16 & 9.164e-16 & 1.265e-15 & 1.819e-15 & 7.871e-14 & 5.598e-13 & 3.131e-12 & 7.056e-12 & 1.398e-11 & 2.613e-11 & 3.811e-11 & 4.963e-11 & 6.053e-11 \\
3 & 1 & 3.662e-17 & 3.625e-17 & 4.124e-17 & 5.054e-17 & 6.567e-17 & 8.971e-17 & 1.279e-16 & 1.894e-16 & 2.891e-16 & 4.508e-16 & 1.987e-14 & 1.476e-13 & 1.016e-12 & 2.676e-12 & 6.112e-12 & 1.291e-11 & 2.005e-11 & 2.717e-11 & 3.406e-11 \\
3 & 0 & 5.680e-18 & 5.697e-18 & 6.488e-18 & 7.930e-18 & 1.021e-17 & 1.372e-17 & 1.918e-17 & 2.772e-17 & 4.115e-17 & 6.197e-17 & 1.786e-15 & 1.396e-14 & 1.326e-13 & 4.246e-13 & 1.164e-12 & 3.023e-12 & 5.476e-12 & 8.346e-12 & 1.144e-11 \\
4 & 3 & 4.033e-16 & 3.707e-16 & 4.219e-16 & 4.811e-16 & 5.567e-16 & 6.781e-16 & 8.705e-16 & 1.144e-15 & 1.536e-15 & 2.103e-15 & 5.222e-14 & 3.610e-13 & 2.252e-12 & 5.502e-12 & 1.169e-11 & 2.323e-11 & 3.498e-11 & 4.630e-11 & 5.674e-11 \\
4 & 2 & 7.546e-17 & 7.361e-17 & 9.554e-17 & 1.142e-16 & 1.364e-16 & 1.774e-16 & 2.514e-16 & 3.682e-16 & 5.525e-16 & 8.418e-16 & 3.153e-14 & 2.262e-13 & 1.486e-12 & 3.758e-12 & 8.163e-12 & 1.624e-11 & 2.421e-11 & 3.179e-11 & 3.879e-11 \\
4 & 1 & 2.871e-17 & 2.853e-17 & 3.229e-17 & 3.836e-17 & 4.831e-17 & 6.451e-17 & 9.013e-17 & 1.309e-16 & 1.956e-16 & 2.977e-16 & 1.101e-14 & 7.980e-14 & 5.566e-13 & 1.491e-12 & 3.481e-12 & 7.586e-12 & 1.209e-11 & 1.666e-11 & 2.107e-11 \\
4 & 0 & 3.995e-18 & 4.021e-18 & 4.456e-18 & 5.319e-18 & 6.833e-18 & 8.506e-18 & 1.161e-17 & 1.648e-17 & 2.416e-17 & 3.621e-17 & 1.315e-15 & 9.893e-15 & 7.532e-14 & 2.202e-13 & 5.838e-13 & 1.543e-12 & 2.877e-12 & 4.477e-12 & 6.200e-12 \\
5 & 4 & 5.545e-16 & 5.101e-16 & 5.345e-16 & 5.980e-16 & 7.035e-16 & 8.219e-16 & 1.011e-15 & 1.278e-15 & 1.663e-15 & 2.230e-15 & 5.837e-14 & 3.916e-13 & 2.265e-12 & 5.339e-12 & 1.116e-11 & 2.220e-11 & 3.355e-11 & 4.428e-11 & 5.376e-11 \\
5 & 3 & 1.282e-16 & 1.278e-16 & 1.420e-16 & 1.690e-16 & 2.149e-16 & 2.637e-16 & 3.532e-16 & 4.925e-16 & 7.125e-16 & 1.063e-15 & 4.172e-14 & 2.906e-13 & 1.746e-12 & 4.176e-12 & 8.704e-12 & 1.692e-11 & 2.502e-11 & 3.254e-11 & 3.912e-11 \\
5 & 2 & 7.653e-17 & 7.476e-17 & 8.187e-17 & 9.513e-17 & 1.153e-16 & 1.496e-16 & 2.000e-16 & 2.790e-16 & 4.050e-16 & 6.076e-16 & 2.524e-14 & 1.789e-13 & 1.100e-12 & 2.686e-12 & 5.718e-12 & 1.129e-11 & 1.677e-11 & 2.185e-11 & 2.632e-11 \\
5 & 1 & 2.262e-17 & 2.159e-17 & 2.333e-17 & 2.676e-17 & 3.186e-17 & 4.143e-17 & 5.466e-17 & 7.517e-17 & 1.077e-16 & 1.603e-16 & 7.189e-15 & 5.332e-14 & 3.471e-13 & 8.958e-13 & 2.070e-12 & 4.606e-12 & 7.511e-12 & 1.049e-11 & 1.328e-11 \\
5 & 0 & 2.629e-18 & 2.455e-18 & 2.621e-18 & 2.974e-18 & 3.497e-18 & 4.469e-18 & 5.769e-18 & 7.731e-18 & 1.079e-17 & 1.568e-17 & 7.253e-16 & 5.670e-15 & 4.002e-14 & 1.127e-13 & 2.974e-13 & 8.075e-13 & 1.548e-12 & 2.445e-12 & 3.394e-12 \\
\tableline
\end{tabular}
\end{center}
\end{tiny}
\end{table}
\end{landscape}

\clearpage

\clearpage

\clearpage

\clearpage

\end{document}